\documentclass[epj]{svjour}

\usepackage{graphics}
\usepackage{amssymb}
\usepackage{amsmath}
\usepackage{xcolor}
\usepackage[ngerman,english]{babel}
\usepackage[switch]{lineno} 
\usepackage{cite}

\begin{document}

\title{Measurements of high-n transitions in intermediate mass kaonic atoms by SIDDHARTA-2 at DA$\mathrm{\Phi}$NE}
\author{F. Sgaramella\inst{1}\thanks{\email {francesco.sgaramella@lnf.infn.it}}
\and M. T\"uchler\inst{2,3}\thanks{\email {marlene.tuechler@oeaw.ac.at}}
\and C. Amsler\inst{2} \and M. Bazzi\inst{1} \and D. Bosnar\inst{4} \and M. Bragadireanu\inst{5} \and M. Cargnelli\inst{2} \and M. Carminati\inst{6,7} \and A. Clozza\inst{1} \and G. Deda\inst{6,7} \and R. Del Grande\inst{8,1} \and L. De Paolis\inst{1} \and L. Fabbietti\inst{8} \and C. Fiorini\inst{6,7} \and I. Fri\v{s}\v{c}i\'c\inst{4} \and C. Guaraldo\inst{1} \and M. Iliescu\inst{1} \and M. Iwasaki\inst{9} \and A. Khreptak\inst{1,10} \and S. Manti\inst{1} \and J. Marton\inst{2} \and M. Miliucci\inst{1} \and P. Moskal\inst{10,11} \and F. Napolitano\inst{1} \and S. Nied\'{z}wiecki\inst{10,11} \and H. Ohnishi\inst{12} \and K. Piscicchia\inst{13,1} \and Y. Sada\inst{12} \and
A. Scordo\inst{1}\thanks{\email {alessandro.scordo@lnf.infn.it}}
\and H. Shi\inst{2} \and M. Silarski\inst{10} \and D. Sirghi\inst{1,5,13} \and F. Sirghi\inst{1,5} \and M. Skurzok\inst{10,11} \and A. Spallone\inst{1} \and K. Toho\inst{12} \and O. Vazquez Doce\inst{1} \and E. Widmann\inst{2} \and C. Yoshida\inst{12} \and J. Zmeskal\inst{2} \and C. Curceanu\inst{1}  
} 
\institute{INFN-LNF, Istituto Nazionale di Fisica Nucleare-Laboratori Nazionali di Frascati, Frascati, 00044 Roma, Italy
\and Stefan-Meyer-Institut f{\"u}r Subatomare Physik, Vienna, 1030, Austria
\and University of Vienna, Vienna Doctoral School in Physics, Vienna, 1090, Austria
\and Department of Physics, Faculty of Science, University of Zagreb, 10000 Zagreb, Croatia 
\and Horia Hulubei National Institute of Physics and Nuclear Engineering IFIN-HH Măgurele, Romania
\and Politecnico di Milano, Dipartimento di Elettronica, Informazione e Bioingegneria, Milano, 20133, Italy
\and INFN Sezione di Milano, 20133, Italy
\and Physik Department E62, Technische Universit{\"a}t M{\"u}nchen, Garching, 85748, Germany
\and RIKEN, Tokyo 351-0198, Japan
\and Faculty of Physics, Astronomy, and Applied Computer Science, Jagiellonian University, Krakow, 30-348, Poland
\and Center for Theranostics, Jagiellonian University, Krakow, Poland
\and Research Center for Electron Photon Science (ELPH), Tohoku University, Sendai, 982-0826, Japan
\and Centro Ricerche Enrico Fermi – Museo Storico della Fisica e Centro Studi e Ricerche “Enrico Fermi”, Roma, 00184, Italy
}
%
%
            
\abstract{
The SIDDHARTA-2 experiment installed at the DA$\mathrm{\Phi}$NE collider of INFN-LNF performed, for the first time, measurements of high-n transitions in intermediate mass kaonic atoms during the data taking campaigns of 2021 and 2022. Kaonic carbon, oxygen, nitrogen and aluminium transitions, which occur in the setup materials, were measured by using the kaons stopped in the gaseous helium target cell with aluminium frames and Kapton walls, and are reported in this paper.\\
These new kaonic atoms measurements add valuable input to the kaonic atoms transitions data base, which is used as a reference for theories and models of the low-energy strong interaction between antikaon and nuclei. Moreover, these results pave the way for future dedicated kaonic atoms measurements through the whole periodic table and to a new era for the antikaon-nuclei studies at low energy.   
}
\maketitle
%
%
\section{Introduction}

\noindent Kaonic atoms represent an ideal tool to study the low-energy regime of Quantum Chromodynamics (QCD) in the strangeness sector, which cannot be described with a perturbative approach. They enable to directly access the K$^-$N interaction at threshold, without the need of an extrapolation as in the case of scattering experiments, since the relative energy between the kaon and the nucleus is already at the level of few keV. This access is crucial to constrain and unify the theoretical descriptions of this interaction.\\
Kaonic atoms have been studied intensely in the 1970s and 1980s, with measurements spanning over a wide range of elements in the periodic table, from lithium to uranium \cite{Wiegand:1971zz,Batty:1979a,Batty:1977,Backenstoss:1973,Backenstoss:1974,Batty:1979b,Barnes:1974,Wiegand:1974,Kunselman:1971,Kunselman:1974,Batty:1981,Cheng:1975}. These measurements serve as database for low-energy antikaon-nuclei studies, constraining the theoretical descriptions of the K$^-$N interaction potential \cite{Batty:1997,Friedman:1994hx}. However, the available data are affected by large experimental uncertainties and several measurements were proven to be at variance with more recent measurements employing modern detector technology, while many more kaonic atoms transitions are not yet measured \cite{Friedman:1994hx}. For these reasons, starting in the late 1990s, a new era of kaonic atoms studies was initiated by the KpX experiment at KEK in Japan, followed by the DEAR and SIDDHARTA experiments at DA$\mathrm{\Phi}$NE, initially motivated by the so-called ``kaonic hydrogen puzzle" \cite{Okada:2007ky,Iwasaki:1997,Bazzi:2011}, which was eventually solved by these experiments. The puzzle consisted of the fact that measurements performed on kaonic hydrogen in the 1970s and 1980s had found an attractive-type interaction \cite{Davies:1979,Izycki:1980,Bird:1983}, in contradiction with the analyses of low-energy scattering data \cite{Martin:1981,Kim:1965,Sakitt:1965}. Similarly, the early data for the kaonic helium-4 $2p$ level shift (see \cite {Wiegand:1971zz,Baird:1983ub}) showed a $>5 \sigma$ discrepancy with theoretical expectations \cite{Batty:1997,Hirenzaki:2000}, the so-called ``kaonic helium puzzle". The puzzle was resolved in 2007 by the measurement of the E570 experiment at KEK \cite{Okada:2007ky}, later confirmed by the SIDDHARTA collaboration \cite{SIDDHARTA:2012rsv}. These examples suggest that the current collection of data from the old experiments is not reliable, and new measurements with higher accuracy and state-of-the-art detectors are required to provide well grounded experimental input to theory.\\
\\
The SIDDHARTA-2 experiment at the DA$\mathrm{\Phi}$NE collider of Laboratori Nazionali di Frascati (INFN-LNF) in Italy is performing high precision kaonic atoms measurements. In 2019, the apparatus was installed on the collider with the main aim to measure the $2p \rightarrow 1s $ transition in kaonic deuterium. During the optimization phase of the collider and of the setup performances in 2021 and 2022, data were collected both with a reduced setup called SIDDHARTINO, which differs in the number of used Silicon Drift Detector (SDD) arrays, and with the full SIDDHARTA-2 setup. To optimize the setup parameters, various measurements were performed with different $^4$He target gas densities \cite{SIDDHARTA2:2022}. 
\\
The target cell consists of an aluminium frame with Kapton foil walls and contains gaseous helium, where the kaons produced by $\phi$-decays are stopped. As a result, multiple kaonic atoms lines are present in the measured X-ray spectra, besides those of kaonic helium. Measurements of these high-n transition energies in the same kaonic atom, combined with the already performed measurements of low-n levels \cite{Friedman:1994hx}, can contribute to separate the one-nucleon interaction from multi-nucleon processes \cite{FRIEDMAN2013170, Wycech:2020vpl}, and lead to a better and more accurate understanding of the kaon-nuclei interaction at low energy.
\\ 
In this work, measurements of several intermediate mass kaonic atoms, such as kaonic carbon, oxygen, nitrogen, and aluminium high-$n$ transition energies are reported, which also represent  the first measurements ever for the reported transitions.

\section{The experimental apparatus}
\begin{figure}
\centering
\resizebox{0.3\textwidth}{!}{
\includegraphics{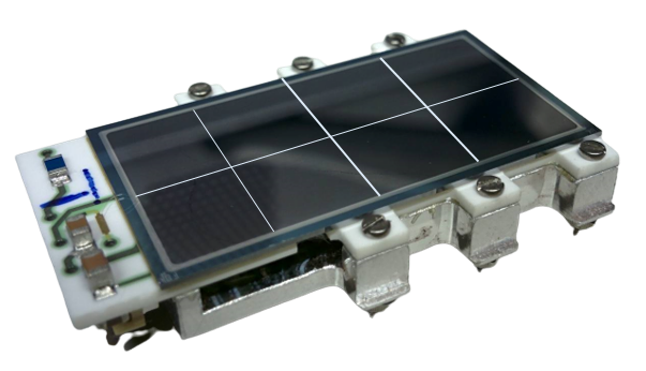}
}
\caption{Silicon drift detector array for the SIDDHARTA-2 experiment with eight read-out cells indicated by the grey lines. The array is mounted on a ceramic carrier.}
\label{fig:sdds}
\end{figure}

\noindent
DA$\mathrm{\Phi}$NE \cite{Milardi:2018sih}, a double ring e$^+$e$^-$ collider at INFN-LNF, provides an ideal environment for precision measurements of kaonic atoms, since it operates at the centre-of-mass energy of 1.02~GeV/$c^2$ and hence produces $\phi$-mesons almost at rest. The charged kaon pairs from $\phi$-decays, produced with a branching ratio of $49.1\%$, are therefore emitted in an almost back-to-back topology with momenta of $\sim$~127~MeV/c and low momentum spread ($\delta p/p \sim 0.1\%$).
The SIDDHARTINO experimental apparatus, installed at the interaction point (IP) of the DA$\mathrm{\Phi}$NE collider, was used for the commissioning of the setup from January 2021 to July 2021, which was followed by the installation of the full SIDDHARTA-2 apparatus and subsequent data taking in April - July 2022. 
\begin{figure*}
\centering
\resizebox{0.5\textwidth}{!}{
\includegraphics{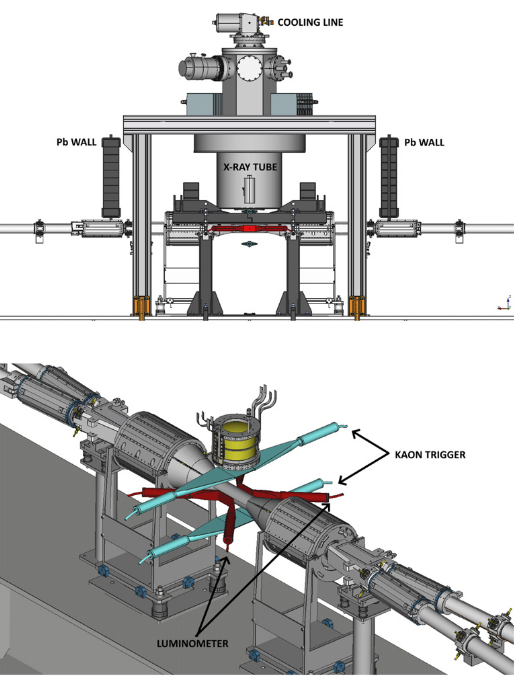}
}
\caption{\textit{Top:} Schematic of the SIDDHARTINO experimental apparatus installed at the DA$\mathrm{\Phi}$NE interaction point \textit{Bottom:} Detailed view of the kaon trigger system (cyan), the luminosity monitor (red) and the target cell (yellow) surrounded by SDDs \cite{SIDDHARTA2:2022}.}
\label{fig:setup}
\end{figure*}
\noindent
A schematic view of the SIDDHARTINO experimental apparatus is shown in the top panel of Fig. \ref{fig:setup}. To suppress the asynchronous background by taking advantage of the back-to-back K$^+$K$^-$ emission, a kaon trigger (KT) system, consisting of two scintillators with photomultiplier (PMT) read-out, is employed, one scintillator being placed below the beam pipe and the other one above, directly in front of the entrance window of the vacuum chamber, as shown in the bottom panel of Fig. \ref{fig:setup}. Only triggered events, where two signals are detected in coincidence by the two scintillators are detected, while the others are discarded. To monitor the beam quality and background in real-time, a luminosity monitor built with plastic scintillators and based on J-PET technology is implemented \cite{Skurzok:2020phi,Moskal:2021medbio,Moskal:2021natcom,Moskal:2021sciad,niedzwicki:2017actphyspol}. The luminometer takes advantage of the difference in the time of flight from the interaction point to the detector, between the Minimum Ionizing Particles (MIPs), originating from e$^+$ and e$^-$ lost by the beams, and kaons. Lead shielding, placed around the vacuum chamber, further suppresses background originating from beam losses and consequent bremsstrahlung. Inside the vacuum chamber, the kaonic atoms are produced inside a cylindrical, lightweight target cell cooled down to $\sim$ 25 K. The target cell has a height of 125~mm and a diameter of 144~mm, and consists of high-purity aluminium bars together with $\sim 150$ $\mu$m thick Kapton sidewalls and a 125 $\mu$m thick Kapton entrance window. \\
\noindent The X-ray detection system surrounds the target cell. It contains large-area Silicon Drift Detector arrays, developed by Fondazione Bruno Kessler (FKB) in Trento, Politecnico di Milano, INFN-LNF and the Stefan Meyer Institute (SMI) in Vienna. Each array features $2 \times 4$ SDD cells with an active area of 0.64 cm$^2$ each. Fig. \ref{fig:sdds} shows one SDD array. For the SIDDHARTINO setup, 8 SDD arrays and hence 64 read-out channels with a total active area of 41 cm$^2$ are employed; for SIDDHARTA-2, 48 arrays are used, with 384 read-out channels and a total active area of 246 cm$^2$. Each SDD signal is amplified by CUBE \cite{5873732}, installed on ceramic carrier structures behind the detectors. The signals are read out and processed by a dedicated ASIC chip called SFERA \cite{Quaglia:2016uox,Schembari:2016IEE}. For in-situ calibration of the SDDs, two X-ray tubes are implemented and periodic calibration runs are performed by the activation of high-purity Ti-Cu foils, mounted on the target cell. The fluorescence lines from these foils are then used for the online calibration of the detectors \cite{Sgaramella:2022}. The SDD system was optimized and characterized in terms of its spectroscopic and timing response, showing a linearity better than 3~eV in the energy range 4 - 14~keV, an energy resolution of ($157.8 \pm 0.3 \pm 0.2$) eV at 6.4~keV \cite{Miliucci:2021wbj} and a long-term stability at the level of 2-3~eV \cite{Sgaramella:2022}, which fulfilled the requirements for the challenging measurements of light kaonic atoms. The final SIDDHARTA-2 setup for the planned kaonic deuterium measurement will employ additional measures for the suppression of background in the form of active, multiple-stage veto systems \cite{revmodphys:2019}, which were not yet implemented during the measurements reported in this work, to enable a direct comparison with the conditions of past kaonic helium experiments \cite{bazzi:2009}.
\section{Data analysis}
\begin{figure*}
\centering
\resizebox{0.8\textwidth}{!}{
\includegraphics{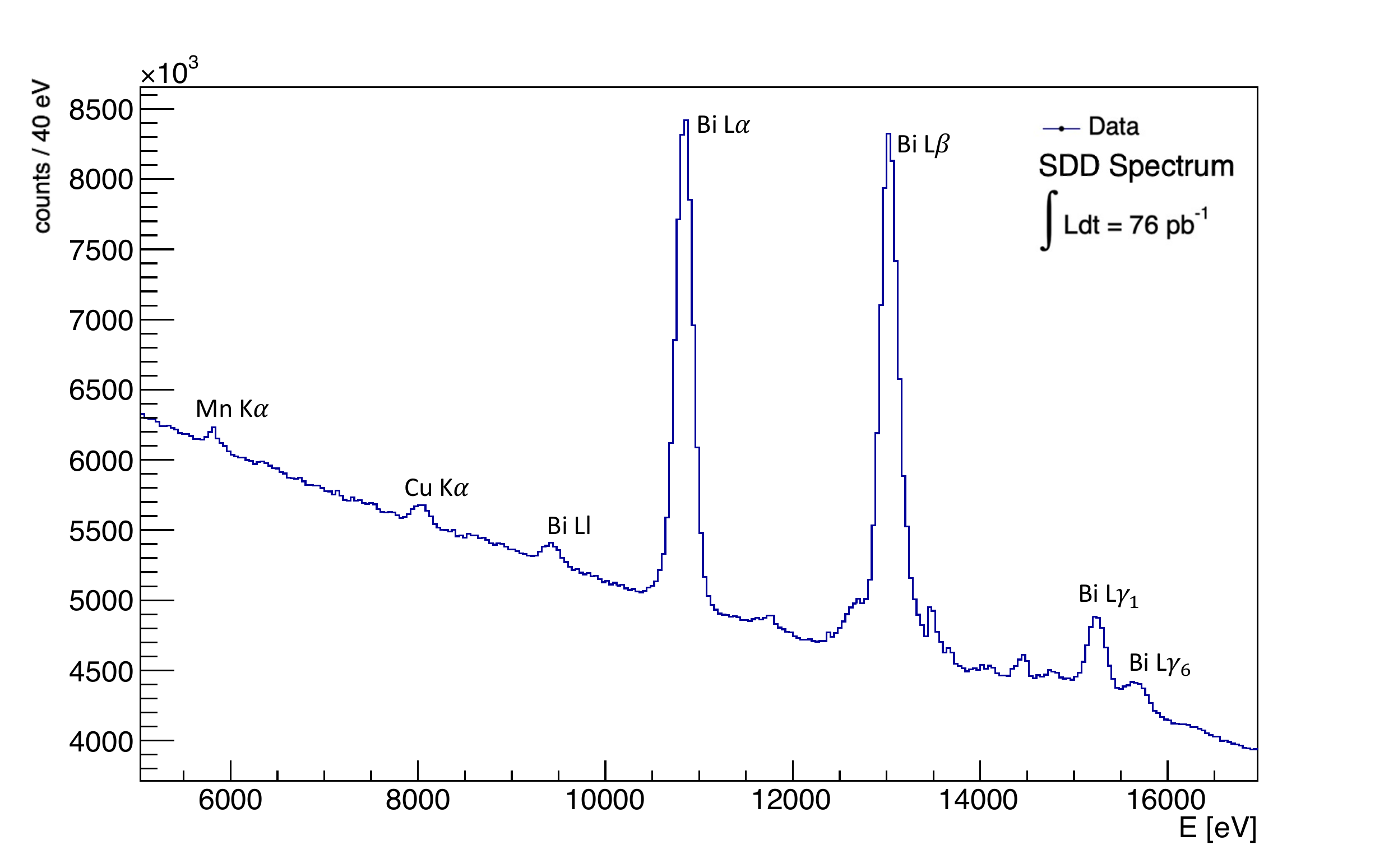}
}
\caption{Inclusive SDD energy spectrum.}
\label{fig:spectrum}
\end{figure*}

\begin{figure}
\centering
\resizebox{0.5\textwidth}{!}{
\includegraphics{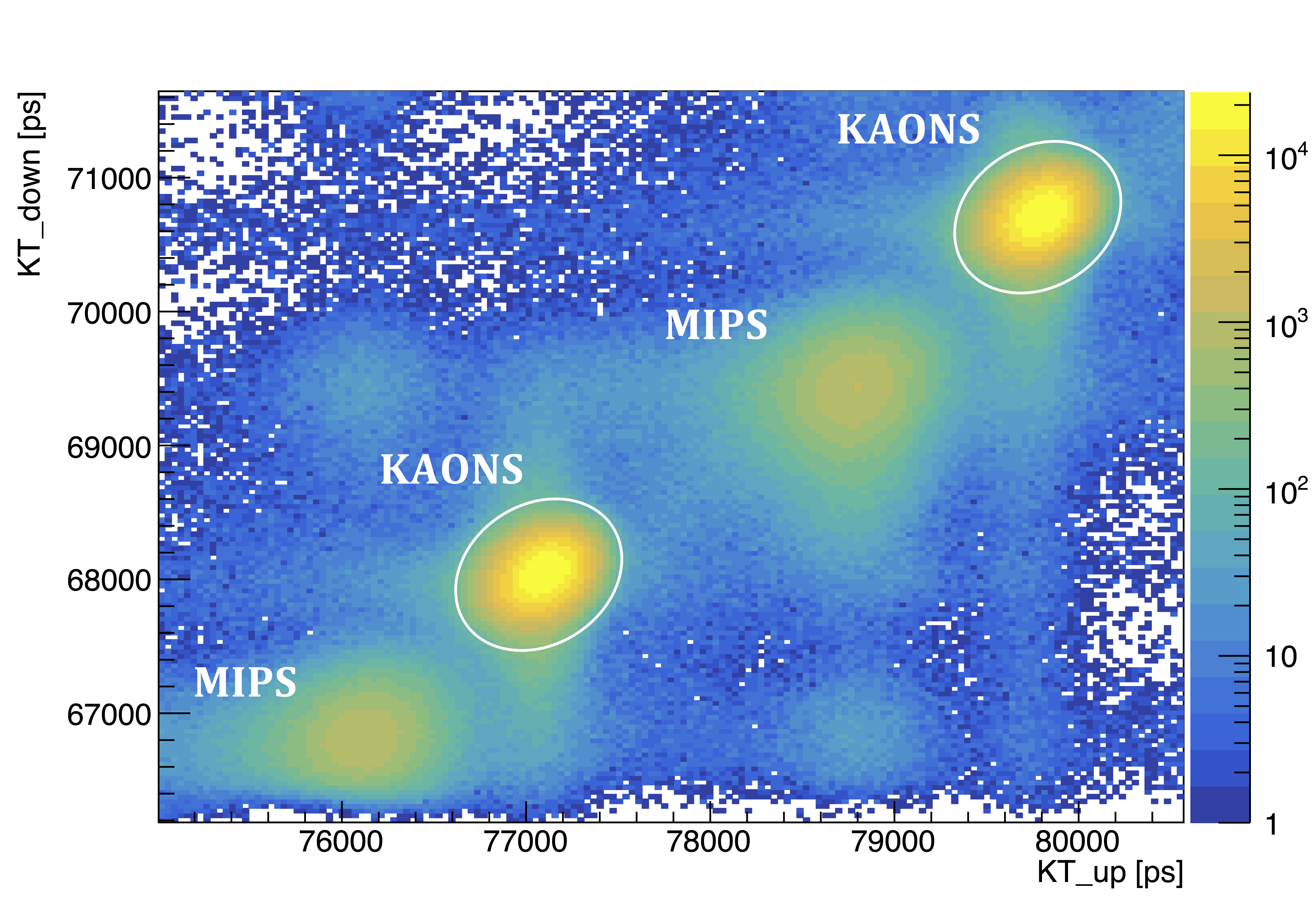}
}
\caption{Scatter plot of the time differences between the top (KT up) and bottom (KT down) scintillators of the kaon trigger and half the DA$\mathrm{\Phi}$NE radio frequency (RF/2). The coincidence events related to kaons (high intensity) are clearly distinguishable from MIPs (low intensity). The double kaon and MIP structures are due to the use of DA$\mathrm{\Phi}$NE RF/2 as time reference (see text).}
\label{fig:KT}
\end{figure}

\begin{figure}
\centering
\resizebox{0.49\textwidth}{!}{
\includegraphics{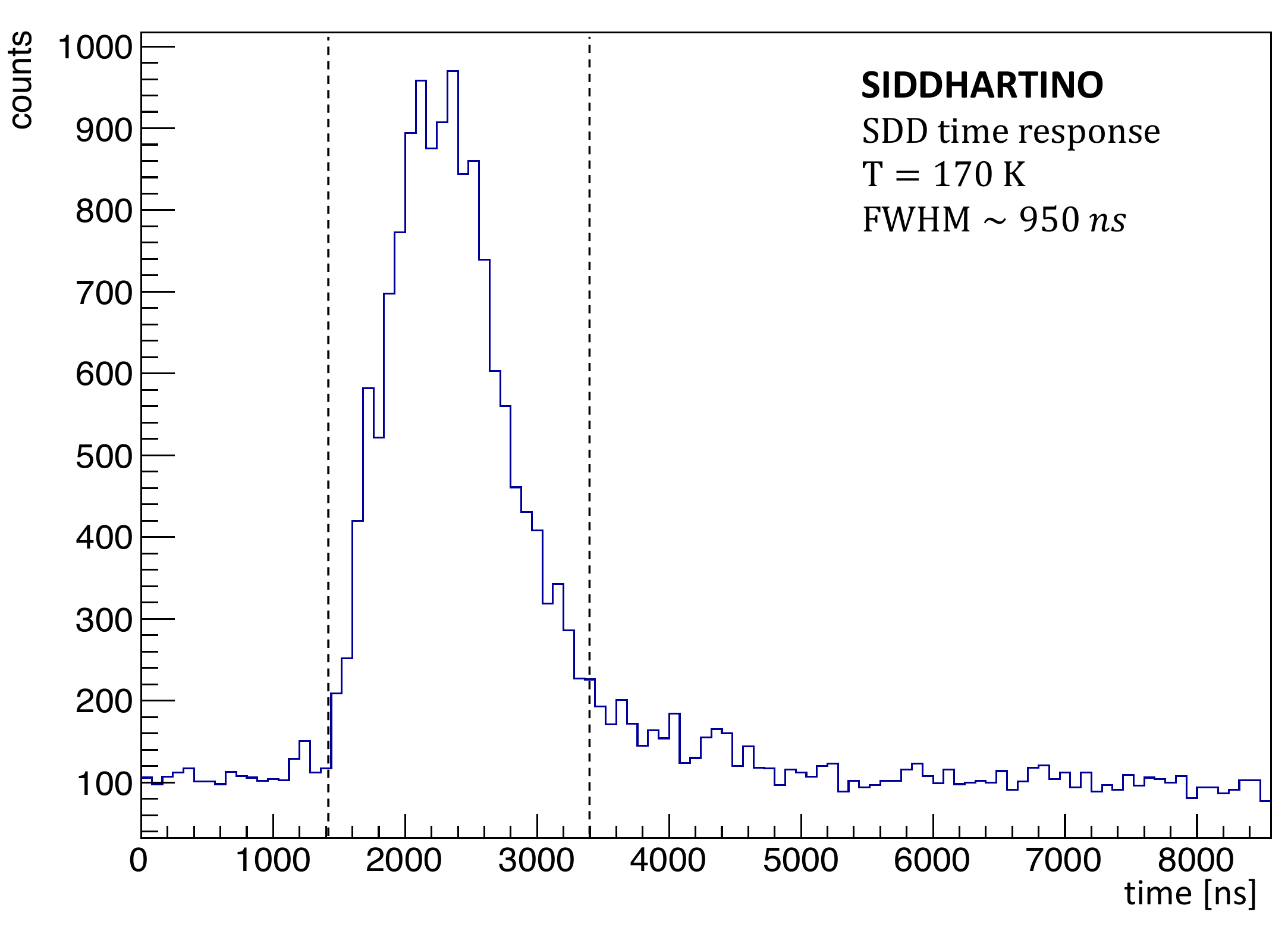}
}
\resizebox{0.49\textwidth}{!}{
\includegraphics{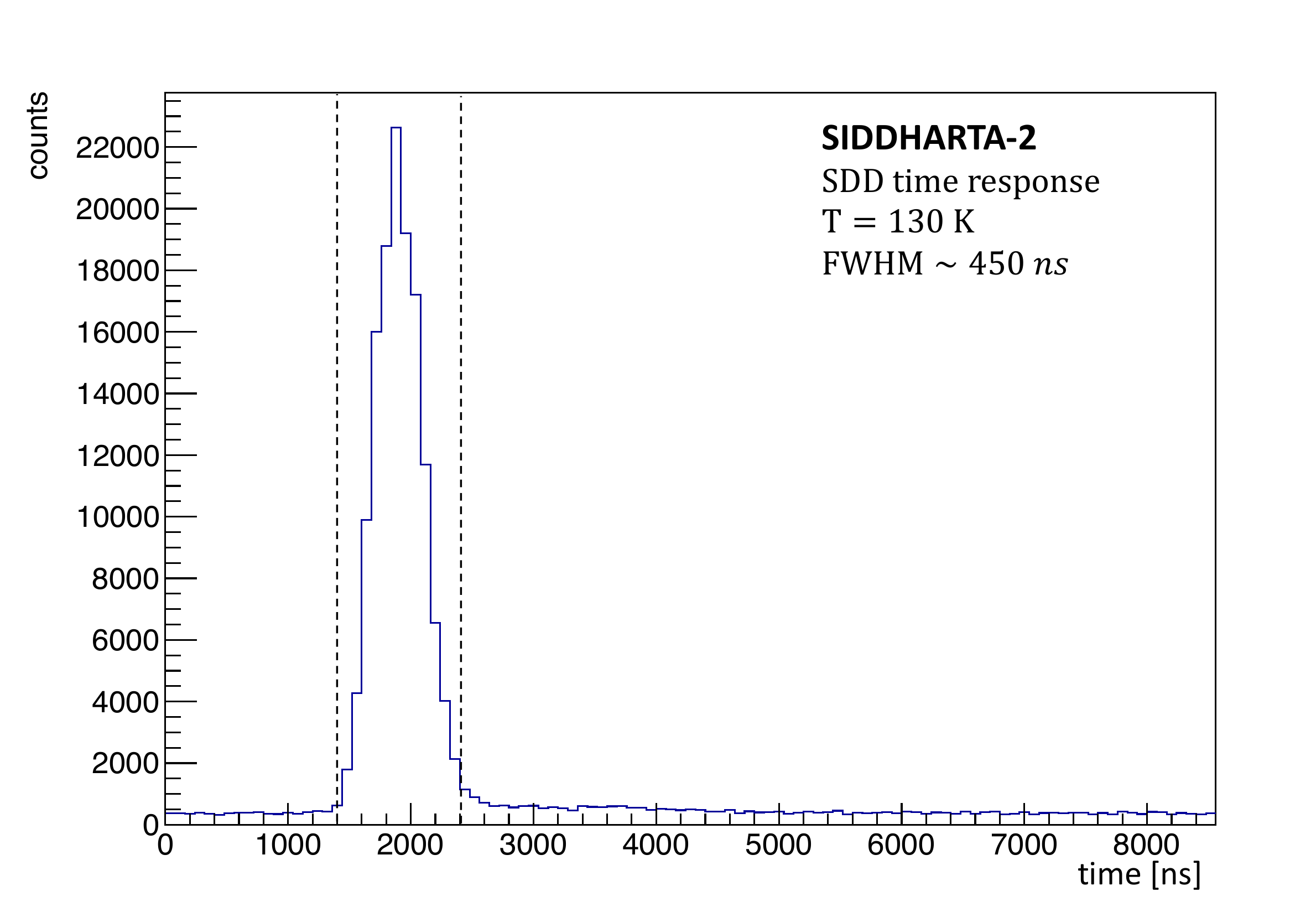}
}
\caption{Distribution of the time difference between the SDD X-rays hit and the trigger signals during the SIDDHARTINO (top), and the SIDDHARTA-2 (bottom) runs. The dashed lines represent the drift time window cut used to reduce the background.}
\label{fig:DrifTime}
\end{figure}

\begin{figure*}
\centering
\resizebox{1\textwidth}{!}{
\includegraphics{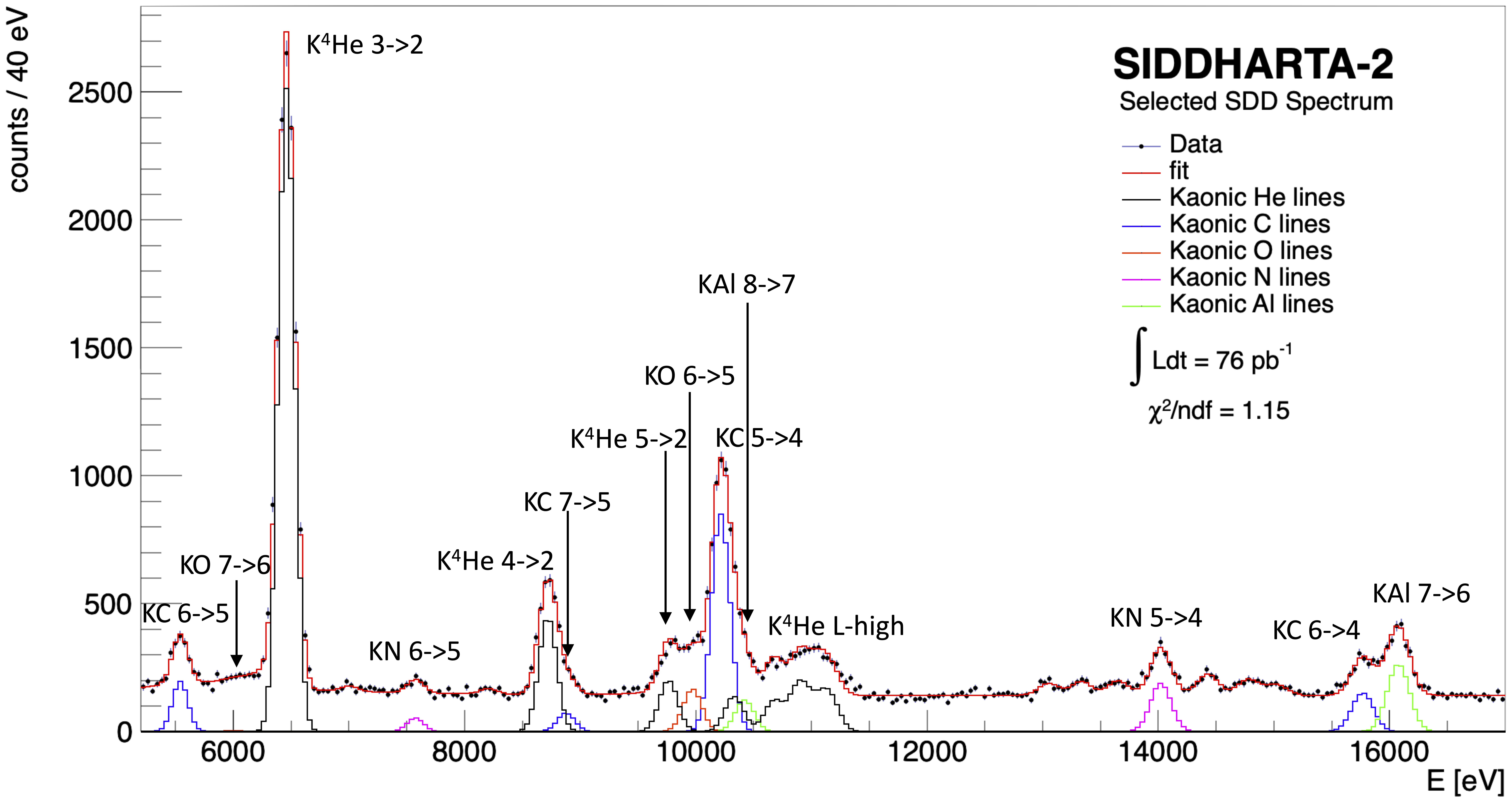}
}
\caption{SDD energy spectrum and fit of SIDDHARTA-2 and SIDDHARTINO summed data after background suppression (see text). The kaonic helium signals are seen as well as the kaonic carbon (KC), oxygen (KO), nitrogen (KN) and aluminium (KAl) peaks.}
\label{fig:fit}
\end{figure*}

\noindent
Data were accumulated with the helium gas target for a total integrated luminosity of 76 pb$^{-1}$. The first run took place with SIDDHARTINO during the beam optimization phase of DA$\mathrm{\Phi}$NE in Spring 2021 and was dedicated to optimize the collider and setup performances. Following this run, a new kaonic helium measurement was performed with the complete SIDDHARTA-2 setup, to assess and adjust the experimental apparatus in view of the kaonic deuterium measurement planned to be performed in 2023.\\
Fig. \ref{fig:spectrum} shows the inclusive energy spectrum obtained by summing all the calibrated data collected with the SDDs. In the spectrum, the fluorescence X-ray transition lines from the activation of the setup materials by the particles lost from the beams are be clearly seen. The Cu K$_\alpha$ line was produced in the calibration foils on the target cell, while the Bi lines originate in the ceramic carriers behind the SDDs.
\noindent
Since most of the background produced by DA$\mathrm{\Phi}$NE is due to particles lost from the beams due to the Touschek effect, and thus is asynchronous with the K$^+$K$^-$ production, it can be drastically reduced by using the kaon trigger. Only the X-ray SDD signals within a 5 $\mu$s time window with respect to the trigger signal can pass the selection (Trigger cut), resulting in a background reduction of a factor 1.3$\cdot 10^4$. The 5 $\mu$s time window was optimized for the front-end electronics which acquires and processes the signals coming from the SDDs. \\
The MIPs, originating from the beam-beam and beam-gas interactions can produce accidental trigger signals when passing through the two scintillators of the kaon trigger in coincidence. The K$^+$K$^-$ pairs detected by the kaon trigger are identified by time of flight (ToF), allowing to distinguish whether a trigger signal is produced by kaons or MIPs. The time difference between the trigger signal and the DA$\mathrm{\Phi}$NE radiofrequency (RF), which provides a reference for each collision, was measured. Fig. \ref{fig:KT} shows the correlation of the mean time for the two scintillators of the kaon trigger. The K$^+$K$^-$ pair events are clearly distinguishable from the MIPs (ToF cut). Since the $\sim$ 370 MHz DA$\mathrm{\Phi}$NE RF cannot be handled by the Constant Fraction Discriminators (CDF) used to process the kaon trigger signals, half the radio frequency (RF/2) was used as reference. Thus, every kaon trigger signal can be associated in time with one of the collisions that occurred over a CDF time period, and consequently the kaons and MIPs appear in two clusters, as shown in Fig. \ref{fig:KT}.
\\Finally, the time information provided by the SDDs was used to further reduce the electromagnetic background. Fig. \ref{fig:DrifTime} shows the time difference between the kaon trigger signal and the X-ray hits on the SDDs for the SIDDHARTINO runs and the SIDDHARTA-2 runs. The peaks represent the X-ray events in coincidence with the kaon trigger, while the flat distribution results from uncorrelated events. The SDDs' timing resolution, which depends on the temperature \cite{Miliucci:2022lvn}, determines the acceptance window (Drift Time cut). For the SIDDHARTINO run, the SDDs' temperature was 170 K and the time window was 950 ns (FWHM) (Fig. \ref{fig:DrifTime} top). In the SIDDHARTA-2 run, the SDDs' temperature was reduced to 130 K, improving the time window to 450 ns (Fig. \ref{fig:DrifTime} bottom). Summing the data of all runs, this resulted in a further reduction of background by a factor two.\\
\noindent
In Fig. \ref{fig:fit}, the X-ray spectrum of the summed data for the SIDDHARTINO and SIDDHARTA-2 runs after application of the event selections in Table \ref{tab:selection} is shown. The kaonic atoms signals are now clearly visible. The peaks highlighted in the figure correspond to the X-ray emissions from kaonic atoms formed in the helium gas and in the components of the target cell. 
\begin{table}
\caption{Data selection steps to reduce the background together with the number of events passing each requirement and the associated background rejection factor}
\label{tab:selection}
	\begin{tabular}{lcc} 
	\hline\noalign{\smallskip}
		 \textbf{Selection} & \textbf{Events} & \textbf{Rejection Factor}\\
	\noalign{\smallskip}\hline\noalign{\smallskip} 
		 No cut         &   $1.6\cdot 10^9$ &   //\\
		 Trigger cut    &   118598 &   1.3$\cdot 10^{4}$\\
		 ToF cut        &   78423  &   1.5\\
		 Drift Time cut &   39163  &   2.0\\
	\noalign{\smallskip}\hline
	\end{tabular}
\end{table}
\begin{table}	
\caption{Kaonic carbon, oxygen, nitrogen and aluminium transition energies from the fit of the data in Fig. \ref{fig:fit}.}
\label{tab:transitions}
	\begin{tabular}{lr} 
	\hline\noalign{\smallskip}
		 \textbf{Transition} & \textbf{Energy (eV)} \\ 
	\noalign{\smallskip}\hline\noalign{\smallskip}
		 K$^-$C (6$\rightarrow$5)     &    5541.7 $\pm$ 3.1 (stat) $\pm$ 2.0 (syst)\\
		 K$^-$C (7$\rightarrow$5)     &   8890.0 $\pm$ 13.0 (stat) $\pm$ 2.0 (syst)\\
		 K$^-$C (5$\rightarrow$4)     &   10216.6 $\pm$ 1.8 (stat) $\pm$ 3.0 (syst)\\
		 K$^-$C (6$\rightarrow$4)     &   15760.3 $\pm$ 4.7 (stat) $\pm$ 12.0 (syst)\\
		  \hline
		 K$^-$O (7$\rightarrow$6)     &    6016.0 $\pm$ 60.0 (stat) $\pm$ 2.0 (syst)\\
		 K$^-$O (6$\rightarrow$5)     &    9968.1 $\pm$ 6.9 (stat) $\pm$ 2.0 (syst)\\
		  \hline
		 K$^-$N (6$\rightarrow$5)     &   7577.0 $\pm$ 17.0 (stat) $\pm$ 2.0 (syst)\\
		 K$^-$N (5$\rightarrow$4)     &   14010.6 $\pm$ 8.2 (stat) $\pm$ 9.0 (syst)\\
		  \hline
		 K$^-$Al (8$\rightarrow$7)    &   10441.0 $\pm$ 8.5 (stat) $\pm$ 3.0 (syst)\\
		 K$^-$Al (7$\rightarrow$6)    &   16083.4 $\pm$ 3.8 (stat) $\pm$ 12.0 (syst)\\
	\noalign{\smallskip}\hline
	\end{tabular}
\end{table}	
\noindent The energy of each kaonic atoms transition was obtained from a fit of the spectrum. The SDDs' energy response for every transition line reported is described by the convolution of a Gaussian with an exponential function to reproduce the incomplete charge collection and electron-hole recombination effect \cite{Miliucci:2022lvn}, except for the kaonic helium lines, which were fitted with Voigt functions to account for the intrinsic linewidth of the transition due to the possible effect of the strong interaction. 
For the other high-n transition lines, the widening induced by strong interaction is known to be negligible \cite{Friedman:1994hx}.\\
In addition, two exponential plus a constant function are used to describe the background shape. The final fit function properly reproduces the data distribution, with a $\chi^2$/ndf = 1.15 in the energy range from 5 keV to 17 keV.\\
Several high-$n$ transition energies in intermediate mass kaonic atoms, such as kaonic carbon, oxygen, nitrogen and aluminium are measured here for the first time. The discussion about kaonic helium transition energies goes beyond the scope of this article and for the SIDDHARTINO data is reported in \cite{SIDDHARTA2:2022}. The kaonic carbon, oxygen and nitrogen transitions are the result of kaons stopped in the Kapton walls, whereas the kaonic aluminium transitions were produced by kaons stopped in the top and bottom frames of the target cell. The final results for the kaonic C, O, N and Al transitions are shown in Table \ref{tab:transitions}. The associated systematic uncertainties take into account the linearity and stability of the SDDs, as well as the effects produced by the energy calibration procedure, following the method described in \cite{Sgaramella:2022,Miliucci:2021wbj}. Thus, various transitions are measured with a statistical precision better than 10 eV, and for the majority of the measurements the systematic uncertainty is below 3 eV.
\section{Conclusions}
\noindent
The SIDDHARTA-2 experiment performed high precision measurements of a series of intermediate mass kaonic atoms transitions, which also represent the first measurements ever for the reported transitions. Kaonic carbon, oxygen, nitrogen and aluminium X-ray transitions in the 5 - 16 keV energy range were measured during the 2021 and 2022 data taking campaigns, by using kaons stopped in the setup materials.\\
These new data enrich the kaonic atoms transitions database, which is used as input and as test-bed for theories and models of kaon-nuclei interactions at low energies, a field which is still far from being fully understood. The new data added by SIDDHARTA-2 can stimulate a revival of the theoretical activity in the field, towards a better understanding of the strong interaction with strangeness and of the role played by multi-nucleon absorption processes \cite{Friedman:1994hx,Wycech:2020vpl}, with implications extending from particle and nuclear physics to astrophysics \cite{revmodphys:2019,Merafina:2020ffb,Curceanu:2020kkg,Drago:2019tbs}.\\
The series of measurements reported in this paper also show the potential of DA$\mathrm{\Phi}$NE and SIDDHARTA-2 technologies to address high precision kaonic atoms measurements along the whole periodic table, within a future program which was already put forward by the scientific community \cite{Curceanu:2021oqj,Curceanu:2021jgq}.

\section*{Acknowledgments}
\noindent
We thank C. Capoccia from LNF-INFN and H. Schneider, L. Stohwasser, and D. Pristauz- Telsnigg from Stefan Meyer-Institut for their fundamental contribution in designing and building the SIDDHARTA-2 setup. We thank as well the DA$\mathrm{\Phi}$NE staff for the excellent working conditions and permanent support. We acknowledge support from the SciMat and qLife Priority Research Areas budget under the program Excellence Initiative—Research University at the Jagiellonian University. Part of this work was supported by the Austrian Science Fund (FWF): Doctoral program No. W1252-N27, as well as [P24756-N20 and P33037-N]; the EXOTICA project of the Minstero degli Affari Esteri e della Cooperazione Internazionale, PO21MO03; the Croatian Science Foundation under the project IP-2018-01-8570; the EU STRONG-2020 project (Grant Agreement No. 824093; the EU Horizon 2020 project under the MSCA (Grant Agreement 754496); the Japan Society for the Promotion of Science JSPS KAKENHI Grant No. JP18H05402; the Polish Ministry of Science and Higher Education grant No. 7150/E-338/M/2018 and the Polish National Agency for Academic Exchange (grant no PPN/BIT/2021/1/00037).
\bibliographystyle{num} 
\bibliography{cas-ref}
\end{document}